\documentclass[prb,amssymb,twocolumn,showpacs,supersriptaddress,floatfix]{revtex4-1}
\usepackage{amssymb}
\usepackage{amsmath}
\usepackage{graphicx}
\usepackage{bm}
\usepackage{caption}
\usepackage{subcaption}
\usepackage{bm}
\usepackage{natbib,hyperref}
\usepackage{hyperref}
\usepackage{color}
\usepackage[titletoc]{appendix}
\hypersetup{colorlinks=true, 
    linkcolor=red,          
    citecolor=magenta,        
    filecolor=gree,      
    urlcolor=blue           
}

\graphicspath{{images/}}

\begin{document}


\title{Oxygen vacancies dynamics  in  redox-based interfaces: Tailoring the memristive response}
\author{Cristian Ferreyra$^{1}$, Wilson Rom\'an Acevedo$^{1}$, Ralph Gay$^{2}$, Diego Rubi$^{1}$}
\affiliation{$^{1}$GIyA and INN, CNEA, Av. Gral Paz 1499 (1650), San Mart\'{\i}n, Buenos Aires, Argentina.\\
$^{2}$CIC nanoGUNE, Tolosa Hiribidea 76, 20018 Donostia-San Sebasti\'an, Spain.}
\author{Mar\'{\i}a Jos\'{e} S\'{a}nchez}
\affiliation{INN, Centro At{\'{o}}mico Bariloche and Instituto Balseiro, 8400 San Carlos de Bariloche, Argentina.}
\date{\today}%
\begin{abstract}

Redox-based  memristive devices are among the alternatives for the next generation of non volatile memories, but also candidates to emulate the behavior of synapses in neuromorphic computing devices. 
It is nowadays well established that the  motion of oxygen vacancies (OV) at the nanoscale is the key mechanism to reversibly switch metal/insulator/metal structures from insulating to conducting, i.e. to accomplish  the resistive switching  effect.

The control of OV dynamics  has a direct  effect on  the resistance changes, and therefore on different figures of memristive devices, such as switching speed, retention, endurance or energy consumption.
 Advances in this direction demand not only experimental techniques that allow for measurements of OV dynamics, but also of theoretical studies that shed light on the involved mechanisms.
Along this goal,  we analize the OV dynamics in redox interfaces formed when an oxidizable metallic electrode is in contact with the insulating oxide.
We show how the transfer of OV  can be manipulated by using different electrical stimuli protocols to optimize device figures such as the ON/OFF ratio or the energy dissipation linked to the writing process.
Analytical expressions for attained  resistance values, including the high and low resistance states are derived in terms of total transferred OV in a nanoscale region of the interface.
Our predictions are  validated with experiments performed in Ti/La$_{1/3}$Ca$_{2/3}$MnO$_{3}$  redox memristive devices.

\end{abstract}

\maketitle

\section{Introduction}
Resistance random access memory (ReRAM) devices have emerged as one of the  main alternatives to current  flash memory technologies. Besides their potential application in the field of non-volatile memories, they have been also tested as logic  devices \cite{borghetti_2010}, and more recently, in the emergent field of neuromorphics \cite{Yu_2017}.
 
The  physical phenomenon behind ReRAMs is the so-called Resistive Switching (RS) effect,  which is the  reversible and 
non-volatile change of the resistance of  a metal/insulator/metal structure upon the application of electrical stimulus \cite{Meijer_2008, Sawa_2008, Waser_2009}. 

The RS has been ubiquitously found in simple and complex  oxides based devices and, in particular, in manganese oxides known as manganites\cite{Liu_2000} . In these  compounds the switching is usually of bipolar type, which requires opposite  polarities for the electrical stimuli to achieve both the SET (high to low resistance, HR$\rightarrow$LR)  and RESET (low to high resistance, LR $\rightarrow$ HR) transitions. 

It was recently shown that when oxidizable metals such as Ti or Al are used as electrodes,   a thin oxide layer (TiO$_x$ or AlO$_x$) is naturally formed at the interface between the metal electrode and the insulating oxide. In particular, in the case of Ti/LCMO(PCMO)\cite{note}, the manganite is spontaneously reduced after the deposition of Ti. This results in a mixed interface TiO$_x$ / LCMO$_{3-x}$ (PCMO$_{3-x}$) in which
the TiO$_x$ layer behaves as a n-type semiconductor and is in contact with the p-type reduced manganite, forming a n-p diode. In these samples, the  RS behavior has been  related to a redox process involving  the transfer of oxygen ions  through the n-p  layer \cite{herpers,hslee}, while the other interfaces behave as ohmic \cite{herpers2,hossein}. We have recently shown that the redox process is activated after the n-p diode is polarized in direct mode, or in inverse mode above breakdown \cite{acevedo_2018}.

The Voltage Enhanced Oxygen Vacancies drift (VEOV) model was originally developed to explain the RS behaviour in single manganites samples \cite{rozen_2010}   and it was further extended to analyse binary oxides  based devices \cite{ghenzi_2013}.
It has been extensively tested in  RS experiments with several devices of the type  M1/Oxide/M2, with M1 and M2 metallic electrodes (like Pt, Au, Cu, Al), and  oxides compounds ranging from manganites (PCMO, LCMO) and cuprates (YBCO), to   binary oxides like TiO$_{2}$ \cite{rozen_2010, ghenzi_2010, marlasca_2011, ghenzi_2012, ghenzi_2013, Rubi_2013}.

Recently, the VEOV model  has been also adapted   to  mimic the  RS behavior in  Ti/LCMO/Pt  samples, where the mixed  TiO$_{x}$/LCMO$_{3-x}$ interface dominates the memristive behavior of the device as a consequence of the redox process already described \cite{acevedo_2018}.

With quite a few exceptions \cite{ghenzi_2012,wang_2015,tang_2016} most of the theoretical studies disregard the connection between OV dynamics and the manipulation  of the attained resistance states. 
The ability to reversibly control the concentration and profile of OV 
 should have  a straightforward impact on the resistance changes, allowing the improvement of the performance of practical devices. This can lead, for example, to the optimization of switching speeds or to the  minimization of the energy consumption for the writing process.
 Advances in this direction demand not only experimental techniques that allow for measurements of OV dynamics \cite{menzel_2011,jang_2017,bao_2018}, but also of theoretical studies that shed light on the involved  mechanisms.
Along this goal, here we perform a systematic analysis of the dynamics of OV in redox interfaces, and their response to different protocols and stimuli. We show the modelling to predict how the electrical stimuli can be manipulated to control OV dynamics and optimize memristive figures such the ON/OFF ratio or the energy consumption linked to the RESET process. 

In addition, we derive analytical  expressions for the attained  resistance values  in terms of  the total amount of OV transferred  along the  interface and as a function of the applied voltage. This  enables the reconstruction of the  R vs V resistance hysteresis switching loops (HSL).

Our  predictions are validated with experiments performed on the Ti/LCMO interface, demonstrating the capability of this kind of simulations to understand the physics related to redox memristive processes, paving the way to optimize the electrical response of practical devices.

\section {The VEOV model revisited for  mixed  redox interfaces}
\label{sec1}

In order to be self contained and to clarify notation, we describe here the main assumptions and equations of the VEOV \cite{rozen_2010,ghenzi_2013} migration model here adapted 
for the study of RS in the mixed  interface   TiO$_{x}$/LCMO$_{3-x}$ \cite{acevedo_2018}.

The  interface  is the active region for the RS behavior, and it  is modelled as a  1D chain of  $N= Nl +  Nr$ total sites, where    $Nl$ sites are associated to the TiO$_{x}$ layer and  $Nr$ sites to the LCMO$_{3-x}$, respectively.
The links   physically represent  small domains of nanoscopic dimensions in both sub-oxides 
with an initial OV concentration that  might correspond to the pristine state (PS).

 We characterize each domain $i$ along the chain by its resistivity $\rho_{i}$ which is a function of the local OV density, $\delta_{i}$.
An universal feature of oxides  is that their resistivity is dramatically affected by the precise oxygen stoichiometry.  LCMO is a complex oxide that behaves as a p-type semiconductor  in which OV disrupt the   Mn-O-Mn bonds with the concomitant  increment of the resistivity.
On the other hand, TiO$_{x}$,  behave as  n-type semiconductor in which oxygen vacancies increment its conductivity. As a consequence,  we adopt for the first $Nl$  domains associated to  the TiO$_{x}$
 the (most simple) relation between resistivity and OV density:   

\begin{equation}  \label{e1}
 \rho^{l}_{i} = {\rho_{0}}^l - A_{i} \delta_{i}.
\end{equation}
where we  define  ${\rho_{0}}^{l} $ as the  residual resistivity  of the left layer for negligible OV concentration ($\delta_{i}=0$).
As the model  description is given in terms of OV,  we conceive  the   TiO$_{x}$ as an OV doped TiO$_{2}$ and  therefore  ${\rho_{0}}^{l} $  corresponds to   the resistivity  of  TiO$_{2}$, (i.e. $x \sim 2$).

On the other hand, as the resistivity of the  LCMO$_{3-x}$ layer increases due to the presence of OV \cite{asamitsu_1997} we define for sites $i=Nl +1, N$:

\begin{equation}  \label{e2}
 \rho^{r}_{i}=  {\rho_{0}}^{r} + B_{i} \delta_{i},
\end{equation}
 being ${\rho_{0}}^{r}$ the residual resistivity of the stoichiometric LCMO.
The coefficients,  $A_{i}$ and $ B_i$  are specific of each layer (oxide) and  can be taken either  as constants  or  smoothly  dependent on the site position, without affecting the qualitative
behaviour of the simulated  results.

The total resistance  along the interface is computed as
$ R = c \sum_{i=1}^{N} \rho_{i}$, with the scale factor taken for simplicity $c\equiv 1$. 
Following Eqs.(\ref{e1},\ref{e2}) we obtain

\begin{eqnarray}\label{tr}
R &=& \sum_{i=1}^{Nl} \rho^{l}_{i}  + \sum_{i=Nl + 1}^{N} \rho^{r}_{i} , \nonumber \\
 &=& R_{s} - \sum_{i=1}^{Nl} A_{i} \delta_{i} + \sum_{i=Nl+ 1}^{N} B_{i} \delta_{i},
\end{eqnarray}
with $R_{s}\equiv Nl \, {\rho_{0}}^{l} + Nr \, {\rho_{0}}^{r}$ the residual resistance of the interface, which is assumed known.

Given an external  stimulus (either a current $ I(t)$ or a voltage $V(t)$) applied to the interface at time t, the OV density at site \textit{i}  is updated for each simulation step according to the  rate probability  $p_{ij} = \delta_i (1-\delta_j) \exp(-V_{\alpha} + \Delta V_{i})$,   for a transfer 
from site \textit{i} to a nearest neighbor \textit{j}= \textit{i} $\pm 1$. Notice that $p_{ij}$  is proportional to the  OV present at site \textit{i}, and to the available concentration at the neighbour site \textit{j} \cite{rozen_2010}. In order to restrict the dynamics of OV to the interface region,  we take 
$p_{01} =p_{10}=p_{N N+1}=p_{N+1 N}=0$.

In the  Arrhenius factor $\exp(-V_{\alpha} + \Delta V_{i})$, $\Delta V_i$ is the 
local potential drop  at site \textit{i} defined as ${\Delta V}_i (t) = V_{i}(t) - V_{i-1}(t)$ with $V_i(t) = I(t) \rho_{i}= V(t) \rho_{i} / R$.
We denote  $V_\alpha$  the activation energy for vacancy diffusion in the absence of external stimulus. All the energy scales are taken in units of the thermal energy $k_{B}T$ and  we consider  $V_\alpha =V_A$, for sites  in the left layer (TiO$_x$), and  $V_\alpha = V_B$ for those in the right layer 
(LCMO$_{3-x}$). 

The numerical implementation starts with  the input of the initial OV  profile along the interface, $\delta_i (0), \forall i=1..N$. 
Different electrical protocols can be employed. According to  standard RS experiments, we chose the  stimulus  $V(t)$ as  a linear ramp  following the cycle $0 \rightarrow V_{m1}  \rightarrow -V_{m2} \rightarrow 0$ a.u.
At each simulation time step $t_k$ 
we compute the local voltage profile $V_i(t_k)$ and the local voltage drops ${\Delta V}_i (t_k)$.  
Employing the probability rates $p_{ij}$ we obtain the transfers between nearest neighboring sites. 
Afterwards the values $\delta_i(t_k)$ are updated to a new set of densities $\delta_i(t_{k+1})$,
with which we compute, at time $t_{k+1}$, the local resistivities  $\rho_i(t_{k+1})$, 
the local voltage drops under the applied voltage $V(t_{k+1})$, and  finally from Eq.(\ref{tr}) the total resistance $R(t_{k+1})$,  to start the next simulation step at $t_{k+1}$.

\begin{figure}[h]
\begin{center}
\includegraphics[width=1\linewidth,clip]{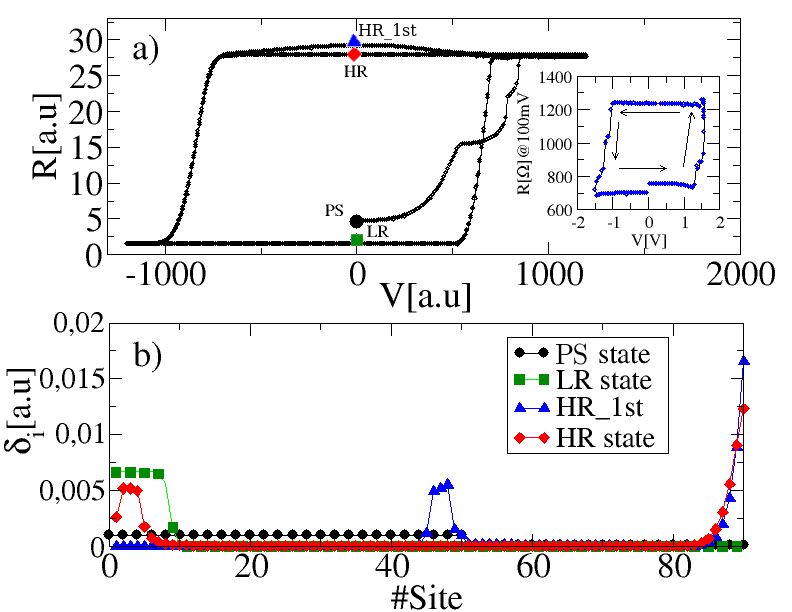}
\end{center}
\caption{a) R vs V (HSL) obtained  within the VEOV model for the 1st and 2nd  cycles of the voltage protocol   $0 \rightarrow V_{m1} \rightarrow -V_{m2}\rightarrow 0$. In the simulations we take $V_{m1}=|V_{m2}|= 1200$ a.u. and $N=90, Nl=50, A=750, B=50, V_A =8.5$, $V_B =6.$ The two  latter values  are chosen following Refs.\onlinecite{ghenzi_2013,nagata_2011}, which report OV diffusions barrier for TiOx (up to 2.5 eV) and LCMO $(\sim $1.3 eV). Inset: Experimental HSL for a single cycle of the voltage protocol, with $V_{m1}=|V_{m2}|= 1.8$V. The arrows indicate the circulation. b) OV density profiles, $\delta_i$, for different resistance states indicated respectively in the  HSL's  of  panel a). See text for details.}
\label{f1}
\end{figure}

The initial configuration of OV in the pristine state (PS) has been taken consistently with the experimentally reported  (low resistance) initial state \cite{herpers,acevedo_2018},  for which  the non-stochiometric TiO$_{x}$ layer ($x<2$)  contributes with a significant conductivity.
The partial oxidation of Ti layer is at  expenses of the reduction of
the  thin LCMO layer that  becomes LCMO$_{3-x}$. This redox process has been clearly identified through spectroscopic characterization by the J\"ulich group in Ref.\onlinecite{herpers}. 
Taking into account this scenario, the initial OV density in the TiO$_x$  is such to grant an appreciable conductivity to this layer. Additionally, as the resistivity of the LCMO$_{3-x}$ increases due to the presence of OV, we  chose  an OV profile  for the PS that  matches these requirements and compatible with the (low resistance) initial state of the complete interface  (see Fig.\ref{f1}(b)).

In Fig.\ref{f1}(a) we show a typical R vs V Hysteresis Switching Loop (HSL), obtained from  the numerical simulations with the VEOV model for  a  symmetric voltage ramp i.e.  $V_{m1}= |V_{m2}|$. Two consecutive cycles are considered in order to show the initial resistance (correspondent to t=0, V=0)  of the PS state, together with the slightly erratic response of the  begining of the  1st voltage cycle.  

In the experiments reported in Refs.\onlinecite{herpers,acevedo_2018}, the RESET process takes place  for positive stimulus and it is  related to the transfer of OV (positive ions) from  the TiO$_x$ layer to the LCMO$_{3-x}$ layer, the first becoming nearly 
stoichiometric  ($x\sim 2$) and thus highly resistive.  At the same time, OV   at the LCMO$_{3-x}$ contribute to increase the resistance. In  Fig.\ref{f1}(b), the OV profile associated to the HR state of the  1st HSL is shown,  in complete agreement with the described behaviour.

The SET transition takes place  for negative stimulus, when the interface returns to a LR state. 
The associated OV profile shown in Fig.\ref{f1}(b) corresponds to the LR state after the completion of the   1st  HSL. In this case OV   accumulate at the left side of the left interface, while for initial  PS, which indeed has a higher resistance value,  the density of OV is constant. 
Besides the initial erratic behavior of the   1st  HSL, the HR and LR states associated to the next cycles  of the voltage protocol become highly repetitive and stable.
As an example, we also show in   Fig.\ref{f1}(b) the OV configuration for the  HR state of the 2nd HSL.

The inset of Fig\ref{f1}(a) shows an experimental HSL recorded for  the Ti/LCMO interface, for a complete cycle of the applied voltage protocol.  The similarity between the simulated and experimental HSL's is remarkable, demonstrating the ability of the  VEOV model to collect the physics of the memristive effect.
Notice that as the amount of transferred OV is controlled by the amplitudes of the electrical stimuli,    different experimental HSL can be obtained by tuning  the voltage (or current) excursions, as it was already discussed  in Ref.\onlinecite{acevedo_2018}, where details of the device fabrication can be  also found.

\section{Resistance  in terms of transferred Oxygen vacancies}

In this section we advance a step further and derive analytical expressions
for the resistance values cast in terms of the  transferred OV as a function of the applied stimulus. 

As in typical experiments, the external electrical stress can be either voltage $V(t)$ or current $I(t)$.
For the sake of simplicity  we consider voltage controlled experiments following the aforementioned protocol, but the  following reasoning will be valid when the  stimulus is $I(t)$.

We start from the initial state, correspondent to the OV configuration depicted in Fig.\ref{f1} b),  consistent with the PS. 
Taking into account  Eq.(\ref{tr}) we write
  \begin{equation}
R(0)=  R_{s} -  A \, al_0 + B \, ar_0,
  \end{equation}
where $R_s$ has been previously defined and we here  denote  the left  and right  initial {\bf areas} (total  number of OV), as $ al_0 \equiv \sum_{i=1}^{Nl} \delta_{i,0}$ and  $ ar_0 \equiv  \sum_{i=Nl+ 1}^{N} \delta_{i,0}$, respectively, with $\delta_{i}(0)\equiv\delta_{i,0}$, the OV density at site $i$ for the initial state. 

Positive voltages  $0 <  V \le V_{m1}$,  move OV (as positive ions) from the  left layer  of the interface (TiO$_x$) to the right layer (LCMO$_{3-x}$), as we have already described. 
For each value of  $V(t) > 0 $ it is possible to compute  the  number of transferred 
vacancies $a^{+}(V(t))$.
Taking into account the conservation of the total number of OV, we define 
$al^{+}(V)= al_{0} - a^{+}(V) $ and $ar^{+} (V) = ar_{0} + a^{+}(V) $. In this way  we  can  write:  
 \begin{eqnarray}\label{ahr}
   R ^{+} (V)& = & R_{s} -  A \, al^{+}(V) + B \, ar^{+}(V)   \; \nonumber \\
   &= & R(0) + \left( A+B \right) a^{+}(V),
 \end{eqnarray}
  showing  that the resistance $R^{+}(V)$ for positive voltages $V(t)$ is  determined by the transferred area $a^{+}(V)$ and  sample specific parameters.
As  $a^{+}$ increases, $R^{+}$ attains higher values and  thus it might be expected that for a   sufficiently strong voltage  $V_{R}\le V_{m1}$,  the RESET transition to the HR state  takes places,  i.e. $ R^{+} (V_R)\equiv$HR.
  
In the next section we will study  the  OV transfer process in order to analyse different scenarios for the  RESET  transition. An important issue that will be addressed  is 
whether the RESET takes place for $a^{+}(V_R)= al_0$ (complete transfer of  the initial number of OV), or alternatively for    $a^{+}(V_R)< al_0$.

For negative voltages,  OV move from the right to left side of the interface.  Defining    
$a^{-} (V)$ as the net transferred area  for a (negative) voltage  $|V| \le V_{m2}$, we can write $al^{-}(V)= al_{0} - a^{+}(V_R) + a^{-} (V) $ and $ar^{-}(V)= ar_{0} + a^{+}(V_R)- a^{-} (V)$, for the left and right interfaces, respectively. 
For simplicity we have  assumed that once the RESET transition takes places for positive polarities and until the reversal of the voltage  polarity, the transfer of vacancies from the right to the left interface is inhibited. This assumption is consistent with the (almost) flat shape of the HSL experimentally  observed for this range of voltages (see  inset of Fig.\ref{f1} a)). Thus we write for   $|V| \le V_{m2}$,  
\begin{eqnarray}\label{alr}
  R^{-} (V) &=&  R_{s} -  A \, al^{-}(V) + B \,  ar^{-}(V)  \; \nonumber \\
   &= &   R_0 +  \left( A+B \right)  \left\lbrace a^{+} (V_R) - a^{-} (V)\right\rbrace .
 \end{eqnarray}
  
In analogy with the previous description, we define the SET transition  for a negative  
voltage $|V_{S}|\le V_{m2}$ with an  associated transferred area $a^{-}(V_S)$. Therefore, from Eq.(\ref{alr}), the low resistance LR state is  $R^{-} (V_{S})=  R_0 + \left( A+B
  \right) \left\lbrace a^{+} (V_R) - a^{-}(V_S) \right\rbrace\equiv$ LR.

We can proceed along for additional cycles of the applied voltage protocol, but  as the  systematics is essentially the same as the one already detailed,  we restrict  the explicit description to  a single cycle.

From Eqs.(\ref{ahr}) and (\ref{alr}) it is possible to  reconstruct  $R$  for a complete cycle of  $V(t)$, i.e. the HSL, once the transferred areas  are determined.

Depending on the relation between  $a^{+} (V_R)$ and  $a^{-}(V_S)$  different scenarios emerge for the LR state. In those cases where $a^{+}(V_R) = a^{-}(V_S)$, the attained LR state results identical to the initial one, see Eq.(\ref{alr}). However in other  cases where 
$ a^{+} (V_R)\lesseqqgtr a^{-}(V_S) $ the  LR $\lesseqqgtr R_0$.
These responses  have been already observed in  the experiment of Ref.\onlinecite{acevedo_2018}   and   give rise to  close or open  HSL after a complete cycle of the voltage excursion.

Besides the  formal simplicity of Eqs.(\ref{ahr}) and (\ref{alr}), the analytical determination of $a^{+} (V)$ and  $a^{-}(V)$ is not a trivial task.
In the following we summarize the main steps followed to obtain   $a^{+}$, and refer the readers to the Appendix for further details.
As  $V(t)$ is a known function of the (discretized) elapsed time $t\equiv \sum_k  t_k $, the total transferred area   can be written as 
 $a^{+}(t)=\sum_k a^{+} ( t_k) $.  To simplify the notation, we denote  $a^{+}_k \equiv a^{+} ( t_k)$. After a lenghtly calculation,  we can write  (see Appendix):
 \begin{equation}
 \label{aplusi}
a^{+}_k = a^{+L}_{k} + a^{+NL}_{k},
\end{equation}
where we  define the linear and non linear contributions respectively as:
\begin{eqnarray}\label{aplusilnl}
a^{+L}_{k}= C_{Nl} \delta_{Nl}(k) \exp(I(k)\rho_{Nl}(k))- C_{Nl+1} \delta_{Nl+1}(k)\nonumber \\
 \exp(-I(k) \rho_{Nl+1}(k)), \nonumber  \\
a^{+NL}_{k}= -\delta_{Nl}(k) \delta_{Nl+1}(k)\lbrace -C_{Nl+1}\exp(-I(k)\rho_{Nl+1}(k)) \nonumber \\
+C_{Nl} \exp(I(k)\rho_{Nl}(k)\rbrace ,\nonumber \\
\end{eqnarray}
with $ I(k)= V(k)/ R(k) $, following the adopted convention.

Notice that in the case of current controlled experiments, in which $I(k)$ is known,
the above equations stress that the transferred area $a^{+}_k$ for the time interval $t_k$ is determined in terms of the density of  OV at the 
{\bf two frontier sites  of the interface}, i.e. $\delta_{Nl} (k)$ and $\delta_{{Nl}+1}(k)$, respectively.
This is a  nontrivial result, that could be experimentally tested using
OV imaging  techniques \cite{menzel_2011} in current controlled experiments, and should contribute to  the design of optimized interfaces for RS experiments.

Equation\eqref{aplusilnl} can  be further simplified taking into account  that  the  activation energies  satisfy $V_A < V_B$,  which implies   $C_{Nl}  >>  C_{Nl+1}$. Taking into account  this approximation, the obtained analytical estimates for $a^{+}_k$  (see Eq.\eqref{wez}) enable the  determination of the  transferred areas as a function of the applied stimulus. 

In the Appendix, we also derived estimates for  $a^{-}_k$ (see Eq.\eqref{wez1}) to compute the  transferred area  $a^{-}= \sum_k a^{-}_k$ for the case of negative applied stimulus.

\begin{figure}[h]
\begin{center}
\includegraphics[width=1\linewidth,clip]{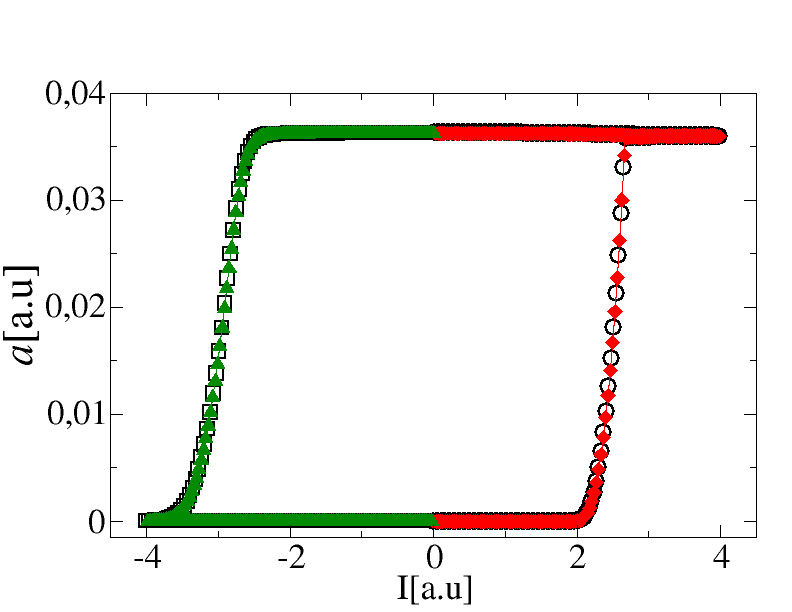}
\end{center}
\caption{Transferred area $a^{+}$(I) ($a^{-}$(I))  for a protocol I(t)= $0 \rightarrow I_{m1} \rightarrow 0$ ($0 \rightarrow  -I_{m2} \rightarrow 0$).
The circles (squares) were  obtained  following the analytical estimates Eq.\eqref{wez}) 
(Eq.\eqref{wez1}). The diamond and triangle symbols  correspond to the numerical calculations employing the VEOV model simulations. We take $I_{m1}= I_{m2}= 4 a.u$} \label{aplusan}
\end{figure}

To give a concrete  example, we consider  protocols controlled by  the current for which the  expressions for   $a^{+}(I)$ and  $a^{-}(I)$ adquire its simplest form,  due to the fact that $I(k)$, the current at each time step  $t_k$, is  known.
Figure \ref{aplusan} shows the analytical estimates for   $a^{+}(I)$ and  $a^{-}(I)$
obtained for a current loop I(t)= $0 \rightarrow I_{m1} \rightarrow -I_{m2}\rightarrow 0$. Notice that the convertion from transferred areas to resistance values is trivial following equations analogous to Eqs.\eqref{ahr} and \eqref{alr}, for the case of current control experiments. Thus, the analytical  reconstruction of the HSL, R vs I,   in terms of the applied stimulus is  fully accomplished.  
The analytical estimates, that only consider the OV at sites $Nl$ and $Nl+1$, result almost indistinguishable from the numerical values (see Fig.\ref{aplusan}) obtained with the VEOV model.  In this  last case  the complete OV profile along the whole interface has to be  updated  at each simulation step  $t_k$, which demands an appreciable computational effort.

An important figure of merit is the HR/LR ratio  which,  from Eqs.\eqref{ahr} and \eqref{alr}, can be expressed as:
\begin{equation}\label{rlh}
\frac{RL -R_0}{RH - R_0}=  \frac{a^{+} (V_R) - a^{-}(V_S)}{a^{+} (V_R)},
 \end{equation}
taking as a reference value $R_0$.

To give a further insight into the transfer area process, in the next section  we  will analyze the dynamics of OV  for different  electrical protocols. This will allow  to determine an optimal stimuli protocol, which shall be confirmed by our experiments on the Ti/LCMO interface.

\section{Dynamics of Oxygen vacancies}

Given the fact that the HR and LR states are essentially determined  by the areas associated to the OV  transferred in the RESET and SET transitions respectively, an interesting  and quite unexplored  aspect is related to the sensitivity  of these processes  to the peculiarities of the  voltage protocol. Along  this goal, in  this section  we analyse the  associated dynamics of OV   for different  applied stimulus. We concentrate in the RESET process  that take place for positive stimulus $V(t)$, but the same analysis can be performed for the SET process.

\begin{figure}[h]
\begin{center}
\includegraphics[width=1.\linewidth,clip]{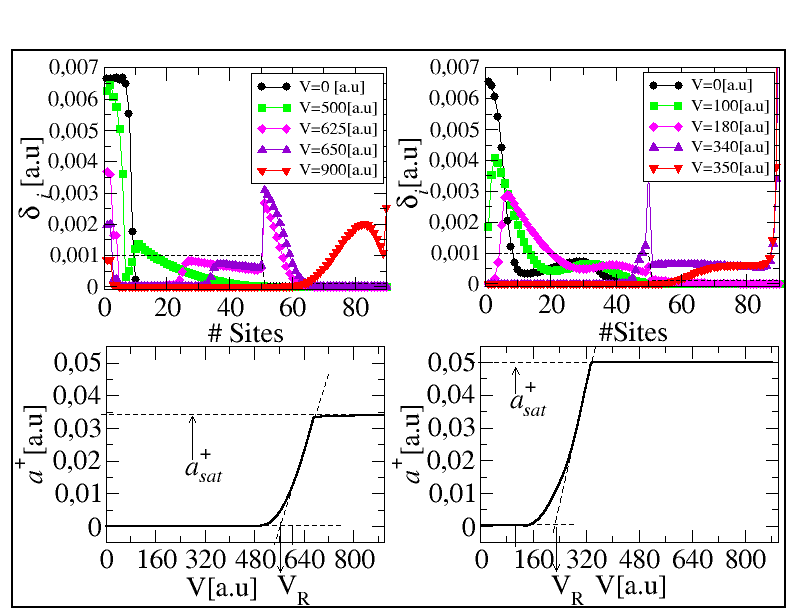}
\end{center}
\caption{Top panels: Spanshots of the OV density profile along the interface for different values of the applied stimulus  V(t), according to two  protocols ramp1(left panel) and ramp2 (right panel), defining a  linear ramp $0 <  V \le V_{m1}$, with  $V_{m1}= 900$ a.u. and  rising time $T_1 = 10$a.u (ramp1) and $T_2 = 225$ a.u.(ramp2), respectively. In dashed black line is the  initial OV profile for $t=0, V=0$,  which gives an  area $al_{0}= 0.05$. Lower panels: transferred area  $a^+$ for different voltage values shown in the upper panels legend.  The RESET voltages $V_R$ are  estimated following the criteria explained in the text.} 
\label{f3}
\end{figure}

The starting point  is the initial OV configuration,  which is shown in both top panels
of  Fig.\ref{f3} labeled by  V=0. This OV distribution defines an initial area 
$al_0$ on the left side of the interface, which we recall corresponds to the TiO$_x$ layer. To  analyse  the  time evolution of this initial  OV density profile,
we consider two   positive voltage excursions  (ramp1 and ramp2) of  a   linear ramp $0 <  V \le V_{m1}$, with  $V_{m1}= 900$ a.u., which differ in the rising time $T_i$ ($T_1 = 10$ a.u and $T_2 = 225$  a.u.), respectively. 

We focus on the evolution of the OV for the  2nd voltage cycle, to  avoid the analysis of   the initial transient in the  OV dynamics, which as we have already described, manifests in an erratic behaviour of the  1st HSL (indeed observed in the experiments).  

In the top panels of Fig.\ref{f3}, we show  snapshots of the density profiles for different values of the voltages which are selected to sample the evolutions. The associated  transferred areas $a^+ (V)$  are shown in the lower panels, respectively.

An  important outcome  is that  the duration of the ramp $T_i$ turns out to be  a  knob that controls whether  the transfer of OV is  complete or not.
Notice that for ramp 1, the transferred area seems to saturate in a value 
$a^+_{sat} \sim 0.035 < al_{0}= 0.05$,  before the completion of the voltage excursion.

This implies that voltage amplitudes  larger than $V\sim 600$ have not  effect in transferring OV from the left to the right side of the interface. In addition, a finite amount of OV  remains in the TiO$_x$ region, consistently with the fact that the complet transfer is not achieved. 

On the other hand, for the ramp 2 (right top panel of Fig.\ref{f3}) the initial area is fully transferred, i.e. $a^+_{sat}= al_{0}= 0.05$. Indeed this is attained for voltage values  lower than $V_{m1}$ (in the present case for $V= 350$, see the OV profile in the right top panel).

From the  plateau in each  plot of  $a^+$ (lower panels) we can  define  a saturated area value,  $a^+_{sat}$.  Doing this we have a  plausible criteria to estimate the reset voltage $V_R$,  as the voltage obtained at  the intersection  between the  horizontal  line correspondent to null transfer area  with the  tangent line  at the  value  ${a^+_{sat}}/2 $.
This is explicitly sketched in both lower panels of Fig.\ref{f3}. The  obtained values of  $V_R$ are  in excellent agreement with  the ones extracted from the HSL in the  VEOV model simulations. 

In the present  example the complete transfer of OV  is attained  for ramp 2, with $T_2 > T_1$. We therefore can conclude that, {\bf for linear  continous ramps}, lower  slopes favour the complete  transfer of OV from the left to right side of the interface, once  the amplitude of the  ramp   $V_{m1}$ exceeds a critial voltage necessary to activate  the transfer. 
From the above analysis  the onset of the RESET transition is clearly identified  with  the ``first arrival" of the OV front  to the right hand side of the interface (LCMO$_{3-x}$).

\begin{figure}[h]
\begin{center}
\includegraphics[width=1.\linewidth,clip]{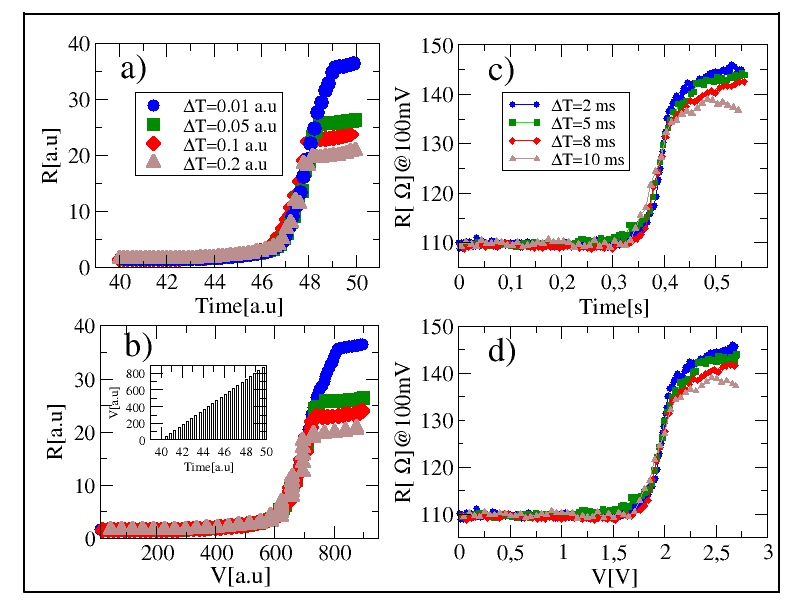}
\end{center}
\caption{Left panels: Simulations with VEOV model. a) R ($R ^{+}$) vs time and b) R($R ^{+}$) vs V. Both plots are for  a series of pulsed protocols following a linear ramp of amplitude $V_m=$ 900 a.u. but differing in the pulse duration $\Delta T$,  as shown in the inset of panel a). A representative ramp is shown in the inset of panel b). 
Right panels: Experiments. c)  Renmant resistance R vs time and  d) R vs V, both according to different experimental protocols differing in the pulse duration.} 
\label{f4}
\end{figure}

Next, we analyze the case of RESET process driven by pulsed voltage ramps, which consist in a series of pulses of increasing amplitude and time width $\Delta T$.  
Consecutive pulses are separated by $\Delta T$ intervals with no applied voltage, as it is shown in the inset  of Fig. \ref{f4} b). This type of  voltage protocol is extensively used in the RS experiments.

We systematically vary $\Delta T$, leaving the total duration of the ramp constant. In this way, shorter $\Delta T$ are associated with ramps with higher number of pulses. 
Figures \ref{f4}(a) and (b) display the corresponding R vs time and R vs V associated to the RESET process, for different $\Delta T$ shown in the legend. 

We recall that  larger  transferred area $a^+ (V)$ implies  larger  renmant resistance  as we deduced in Eq.\eqref{ahr}. 
As it can be observed from the figure, the transferred area is maximized for the shortest pulses, indicating that the OFF/ON  (HR/LR) ratio  is optimized by accumulating a higher amount of short pulses rather than a lower amount of wider ones. 
This non trivial result is experimentally confirmed for the Ti/LCMO interface, as shown in Figs \ref{f4}(c) and (d), which display several RESET processes for voltage pulsed ramps with a fixed total duration of 2.88 s and different $\Delta T$, ranging between 2-10 ms. 
It is  evident that a higher HR final state is achieved for shorter (2ms) pulses, confirming the prediction derived from the simulations. In addition the qualitative agreement with the numerical predictions is remarkable.

Finally, we  address the study of  the RESET process for trains of rectangular  pulses  differing in the time-widths $\Delta T$ and amplitudes $V_0$, but keeping  the product $ V_0 \times {\Delta T} = cte$. 
We start by an  OV profile  defining an initial area $a_{l0}= 0.05$. The RESET process is considered as completed when the initial area is fully transferred (we choose the amplitudes $V_0$ in order to satisfy this requirement).

\begin{figure}[h]
\begin{center}
\includegraphics[width=1\linewidth,clip]{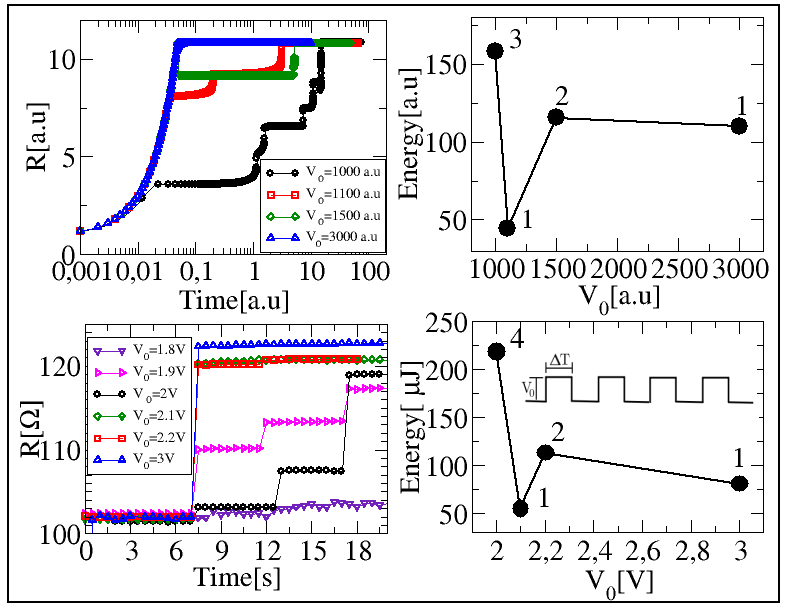}
\end{center}
\caption{Top panels: Simulations. a) R vs  time for different  trains of rectangular pulses
satisfying $ V_0 \times {\Delta T} = cte$. See text for details. b) Electrical energy injected into the system for a complete RESET process. 
Lower panels: Experiment: c) R vs time for different trains of pulses indicated in the legend. d)Electrical energy injected into the system for a complete RESET process.
Panels b) and d) have indexed the number of pulses needed for a complete RESET.}
\label{f5}
\end{figure}

The simulated evolution of  R vs time, for the different trains is displayed in Fig.\ref{f5}(a). It is found that the number of pulses needed to achieve the RESET changes in a non-monotonic way with the amplitude $V_0$. This information is indexed in Fig. \ref{f5}(b), which also displays the electrical energy necessary to complete the RESET process as a function of  $V_0$ for different number of pulses labeled by the numbers. These energies were calculated  as $U=V_0^2 \sum_{i} {\Delta T}_i/ {R_i}$, with $R_i$ is the attained resistance value after the application of the  $i-th$ pulse.  

From this analysis we conclude  that   there is $\{\Delta T$, $V_0\}$ pair  which minimizes the RESET energy. This was indeed verified experimentally for the Ti/LCMO interface, as it is shown in Figs. \ref{f5}(c) and (d). Pulse trains with $V_0$ and  $\Delta T$ ranging between [1.8-3]V  and [1-1.75 ]ms were tested with  the product 
${\Delta T}\times V_0=$3 V ms. 
In the experiment we consider the  RESET process  as completed, when the relative resistance change after the last applied pulse is below 5$\%$. 
Again, the number of pulses necessary to complete the RESET process display a non-monotonic dependence with $ V_0$. To estimate the injected energy during each pulse, we assumed that the resistance increases linearly to its final value during the application of the pulse.

Figure \ref{f5} d) mimics remarkably well the simulated data in Fig.\ref{f5} b). It is found that the RESET energy is minimized for $\Delta T=$ 1.4ms and $V_0 =$2.1V, being necessary in this case a single pulse to complete the process. For higher voltages, the  RESET process can be achieved also with a single pulse but higher energy is required. For lower voltages, it is necessary to accumulate several pulses and therefore the final energy increases. 

\setlength{\parskip}{3mm}

The present analysis clearly shows that the VEOV simulations appear as a powerful tool to analyze oxygen vacancies dynamics in redox memristive systems and predict optimum writing protocols to increase the efficieny of practical devices.

\section{Conclusions}

In summary, we have thoroughly addressed the OV dynamics in redox p-n interfaces by using un updated version of the  VEOV model. The simulations allow to predict the optimum write protocols to control and enlarge the ON/OFF ratio. 
Our results  are  relevant not only for memories optimization, but also for neuromorphic computing applications, as the presence of multilevel resistance states allows mimicking the adaptable synaptic weight of brain synapses \cite{kuzum2013}.

In addtion we found   the optimum stimuli protocol that minimizes the energy consumption linked to the RESET process. This is also important for the optimization of neuromorphic computing devices aiming to emulate the highly efficient energy comsumption of biological systems \cite{merolla2014}.

The  numerical predictions    were fully validated with experiments on the Ti/LCMO memristive interface, demonstrating  the power of this type of phenomenological modelling to predict and optimize the behavior of practical memristive devices.

\section*{Acknowlegments}
We acknowledge support from CNEA, UNCuyo (06/C455) and ANPCyT (PICT2014-1382, PICT2016-0867). 
We acknowledge  L. Hueso who allowed accessing the  nanoGUNE nanofabrication facilities and P. Levy for supporting this collaboration. 

\appendix
\begin{appendices}
  \section{Analytical estimates for the transferred areas}

In this Appendix we derive  the  expressions for the transferred area
$a^{+}_{k}$ from the left to the right side of the interface, in terms of the linear and nonlinear contributions, Eq.\eqref{aplusilnl}. 

We start by defining the rate of OV variation between  neighbours sites $i-1$, $i$ and $i+1$  as:

\begin{equation}\label{DELTA1}
\begin{split}
\Delta_{i}(k)=& (p_{i-1,i}(k)+p_{i+1,i}(k))- \\
& -(p_{i,i-1}(k)+p_{i,i+1}(k)),
\end{split}
\end{equation}
with
\begin{equation}\label{aplus}
 p_{i,j}(k)=C_{i}\delta_{i}(k)(1-\delta_{j}(k))\exp(I(k)\rho_{i}(k))
\end{equation}
and  $C_{i}= \exp{(-V_{\alpha})} $, already introduced  in Sec.\ref{sec1}  of the main text. 
In the following we consider $C_{Nl}= \exp{(-V_A)}$ and $C_{Nl+1}= \exp{-(V_B)}$.

Employing Eq.\eqref{DELTA1}, we write the transferred area (total number of transferred OV) $a_k^{+}$ as: 

\begin{equation}\label{aplus3}
\begin{split}
 a^{+}_{k} \equiv &\sum_{i={N{l}+1}}^{N}\Delta_ i(k)=\sum_{i={N{l}+1}}^{N}\Delta_i^{L}(k)+\sum_{i={N{l}+1}}^{N}\Delta _{i}^{NL}(k)=\\
 &= a^{+L}_{k}+a^{+NL}_{k} ,
 \end{split}
  \end{equation}
 
where we have defined:

\begin{equation}\label{DELTALin}
\begin{split}
\Delta_{i}^{L}(k)=& C_{i-1}\delta_{i-1}(k)\exp(I(k)\rho_{i-1})+\\
&+C_{i+1}\delta_{i+1}(k)\exp(-I(k)\rho_{i+1})-\\
&-C_{i}[\delta_{i}(k)\exp(-I(k)\rho_{i})+\delta_{i}(k)\exp(I(k)\rho_{i})]
\end{split}
\end{equation}
and 
  
\begin{equation}\label{DELTANLin}
\begin{split}
\Delta{i}^{NL}(k)=&-C_{i-1}\delta_{i-1}(k)\delta_{i} \exp(I(k)\rho_{i-1})-\\
&-C_{i+1}\delta_{i+1}(k)\delta_{i}(k)\exp(-I(k)\rho_{i+1})\\
&+C_{i}[\delta_{i}(k)\delta_{i-1}(k)\exp(-I(k)\rho_{i})+\\
&+\delta_{i}(k)\delta_{i+1}(k)\exp(I(k)\rho_{i})].
\end{split}
\end{equation}

Performing the sumations in Eq.\eqref{aplus3} and accounting for the boundary condition $C_{N+1}=0$, we get

\begin{equation}\label{aplusdef}
\begin{split}
\sum_{i=N{l}+1}^{N}\!\Delta_i^{L}(k)=& C_{N{l}}\delta_{N{l}}(k)\exp(I(k)\rho_{N{l}})-\\
&-C_{N{l}+1}\delta_{N{l}+1}(k)\exp(-I(k)\rho_{N{l}+1})
\end{split}
\end{equation}

and 
\begin{equation}\label{aplusdef1}
\begin{split}
\sum_{i=N{l}+1}^{N}\Delta_{i}^{NL}(k)=&-\delta_{N{l}}(k) \delta_{N{l}+1}(k)(C_{N_{l}}\exp(I(k)\rho_{N_{l}})- \\
&-C_{N{l}+1}\exp(-I(k)\rho_{N{l}+1})).
\end{split}
\end{equation}

The linear term $a^{+L}_{k}$ has been written as the sum of two contributions, 
\begin{equation} \label{alapen}
a^{+L}_{k} \equiv P(k) - Q(k),
\end{equation}
 defined as:

\begin{equation}
\label{pq}
\begin{split}
P(k)= &C_{Nl}\delta_{Nl}(k) \exp( I(k)\rho_{Nl}(k)), \\
Q(k)=& C_{Nl+1}\delta_{Nl+1}(k) \exp(-I(k)\rho_{Nl+1}(k)).
\end{split}
\end{equation}

Analogously, we write the nonlinear term as:
\begin{equation}
a_{k}^{+NL}\equiv S(k)-T(k),
\end{equation}
with
\begin{equation}\label{st}
\begin{split}
S(k)=& C_{Nl+1}\delta_{Nl}(k)\delta_{Nl+1}(k)\exp(-I(k)\rho_{Nl+1}(k)) , \\
T(k)=& C_{Nl}\delta_{Nl}(k) \delta_{Nl+1}(k)\exp(I(k) \rho_{Nl}(k))).
\end{split}
\end{equation}

Notice that for  current controlled experiments in which the current $I(k)$ is known by input, the  tansferred areas at each time interval $t_k$ only depend on OV densities and resistivities at the sites $Nl$ and $Nl+1$, respectively. 

As we already mentioned in Sec.\ref{sec1}, the activation energies for OV diffusion
satisfy $V_A < V_B$ and thus   $C_{Nl+1}<<C_{Nl}$. Therefore, we can safely approximate:
\begin{equation}\label{wez}
a_k^{+}=a_k^{+L}+a_k^{+NL}\approx P(k)-T(k).
\end{equation}

This equation can be updated for the next time interval $t_{k+1}$ employing 

\begin{equation}\label{etaplus1}
\delta_{i}(k+1)=\delta_{i}(k)+\Delta_{i}(k),
\end{equation}
for  $i= Nl-1$, $Nl$ and  $Nl+1$, respectively.  

From Eq.\eqref{aplus} and Eq.\eqref{DELTA1} we write after a straightforward algebra:

\begin{equation}
\begin{split}
\Delta_{Nl}(k)&=C_{Nl}\delta_{Nl}(k)[-2\cosh(I(k)\rho_{l}(k))+\\
&+\delta_{Nl+1}(k)\exp(I(k)\rho_{Nl}(k))+\\
&+\delta_{Nl-1}(k)\exp(-I(k)\rho_{Nl}(k))]+\\
&+(1-\delta_{Nl}(k))\\
&[C_{Nl+1}\delta_{Nl+1}(k)\exp(-I(k)\rho_{Nl+1}(k))+\\
&+C_{Nl-1}\delta_{Nl-1}(k)\exp(I(k)\rho_{Nl-1}(k))].
\end{split}
\end{equation}

Performing the sustitution $k\rightarrow k+1 $, replacing Eq.\eqref{etaplus1} in   $P(k)$ (Eq.\eqref{pq}), and taking into account Eq.\eqref{e1} in the main text,  we obtain:

\begin{equation}
\begin{split}
P(k+1)=&C_{Nl}[\delta_{Nl}(k)+\Delta_{Nl}(k)]\\
&\exp\bigl( I(k+1)(\rho_0^{l}-A( \delta_{Nl}(k)+\Delta_{Nl}(k)\bigr).
\end{split}
\end{equation}

In  a similar way  we derive, after updating $T(k)$ in Eq.\eqref{st}, 
\begin{equation}
\begin{split}
T(k+1)=&C_{Nl}[\delta_{Nl}(k)+\Delta_{Nl}(k)][(\delta_{Nl+1}(k)+\Delta_{Nl+1}(k)]\\
&\exp\bigl(I(k+1)(\rho_0^{l}-A ( \delta_{Nl}(k)+\Delta_{Nl}(k))\bigr).
\end{split}
\end{equation}

Employing these  two  last equations we compute $a_{k+1}^{+} \approx P(k+1)-T(k+1)$.

Following the described prescription iteratively,  the total transferred area
$a^{+}$ after an elapsed time $t_{+}=\sum_k t_k$, can be computed under the present assumptions. 

The same procedure can be applied to compute $a^{-}$ for negative electrical estimulus. 
Assuming that the current protocol $0\rightarrow I_{m1} \rightarrow 0$ is completed for a time 
$T_{+}=\sum_k^{K_+} t_k $,  the initial condition for the negative current protocol $0\rightarrow -I_{m2} \rightarrow 0$ should be taken as the  OV profile at time $T_+$, i.e. $\delta_{i}(K_+)$.

 To avoid repetition we  give  below the final expression, valid for $k>K_+ $:
 
\begin{equation}\label{wez1}
\begin{split}
a_k^-=&C_{N_{l}}\delta_{N_{l}}(k)\exp(I(k)\rho_{N_{l}})-\\
&-C_{N_{l}+1}\delta_{N_{l}+1}(k)\exp(-I(k)\rho_{N_{l}+1})+\\
&+\delta_{N_{l}}(k) \delta_{N_{l}+1}(k)(C_{N_{l}+1}\exp(-I(k)\rho_{N_{l}+1})-\\
&-C_{N_{l}}\exp(I(k))\rho_{N_{l}})).
\end{split}
\end {equation}

\end{appendices}

\end{document}